# Accurate parameter estimation using scan-specific unsupervised deep learning for relaxometry and MR fingerprinting

*Mengze Gao[1], Huihui Ye[2], Tae Hyung Kim[3,4], Zijing Zhang[2], Seohee So[5], Berkin Bilgic[3,4]*

[1] Department of Precision Instrument, Tsinghua University, Beijing, China
[2] State Key Laboratory of Modern Optical Instrumentation, College of Optical Science and Engineering, Zhejiang University, Hangzhou, China
[3] Harvard Medical School, Boston, MA, USA
[4] Athinoula A. Martinos Center for Biomedical Imaging, Charlestown, MA, USA
[5] School of Electrical Engineering, Korea Advanced Institute of Science and Technology, Daejeon, Republic of Korea

## Synopsis

**We propose an unsupervised convolutional neural network (CNN) for relaxation parameter estimation. This network incorporates signal relaxation and Bloch simulations while taking advantage of residual learning and spatial relations across neighboring voxels. Quantification accuracy and robustness to noise are shown to be significantly improved compared to standard parameter estimation methods in numerical simulations and in vivo data for multi-echo $T_2$ and $T_2^*$ mapping. The combination of the proposed network with subspace modeling and MR fingerprinting (MRF) from highly undersampled data permits high-quality $T_1$ and $T_2$ mapping.**

## Introduction

Parameter estimation methods e.g. dictionary-matching (1), variable projection (Varpro) (2), and least-squares solvers estimate relaxation parameters on a voxel-by-voxel basis, but fail to exploit spatial relations between voxels, resulting in poor estimates in acquisitions with low signal differentiation or high acceleration.

Prior work on deep learning-based relaxation parameter quantification aimed to learn a direct mapping from contrast-weighted images to parameter maps (3–6), which required large datasets and high-quality reference parameter maps. Such pre-trained networks may fail to generalize to new protocols with different parameter settings and to the presence of out-of-distribution features e.g. pathology.

A recent unsupervised network eliminated the need for high-quality reference maps incorporating the Bloch signal model into the network's loss function (7). This reduced the computational burden of parameter estimation in CEST, and relied on a dataset consisting of multiple subjects.

Herein, we propose an unsupervised convolutional neural network (CNN) for parameter estimation by learning a mapping from contrast-weighted images to the desired parameter maps. This is flexible enough to incorporate Bloch- as well as subspace-modeling in the loss function, and operates in a scan-specific manner by only using data from an individual subject.

## **Methods**

In vivo data were acquired using a Siemens Prisma system with 32ch reception.

$T_2^*$ mapping   Multi-echo 2D gradient echo (2D-GRE) acquisition was performed on a volunteer with matrix size = 224×210×58, 10 echo times ($TE_1$ = 6ms, $\Delta TE$ = 6ms) and TR = 5s.

$T_2$ mapping   Single-echo spin-echo acquisitions were made on a volunteer with 4 echo times ($TE_1$ = 43ms, $\Delta TE$ = 24ms), matrix size = 348×384×58 and TR = 7s.

Numerical simulation   To test the robustness to noise, $M_0$ and $T_2$ data were generated using the output of the network, which were then used to synthesize multi-echo images. Gaussian noise with variance = 0.001 was added to these multi-echo data to generate noisy inputs.

MRF acquisition   A 2D FISP sequence (8) was used to acquire ground-truth MRF data. The acquisition was repeated 6-times to collect 6 interleaves to alleviate aliasing. 30 slices with 4 mm thickness at matrix size = 220×220 required 6 min/interleaf. Data was reconstructed using sliding window processing (9) with a window of size = 6. Parameter maps from this 6-interleaf data were used as references. The proposed reconstruction used only a single interleaf's data.

Network structure [Fig1]:   A combination of convolutional, instance normalization and activation layers (ReLU) are used in the beginning, followed by nine residual blocks. SSIM loss was employed in $T_2$ and $T_2^*$ mapping, and L1 was used for MRF.

Synthesizing multi-echo data for $T_2$ (and $T_2^*$) mapping   The output $M_0$ and $T_2$ maps from the network were used for synthesizing multi-echo image $\rho_{TE}$ due to: $\rho_{TE} = M_0 \times \exp\left(-TE/T_2\right)$.

Synthesizing time-series data for MRF   To circumvent the utilization of a non-differentiable Bloch dictionary-matching, we employed a subspace approach where the MRF dictionary was compressed to 6 coefficients (to reach 95% power in singular values). The output of the network was these coefficient maps c, which was multiplied with the subspace matrix $\phi$ (size = 6×600) to obtain the synthetic MRF time-series $\rho_{TR}$ due to $\rho_{TR} = c \times \phi$.

## Results

**Fig2** shows the estimated parameter maps after adding Gaussian noise to the input echoes. The proposed unsupervised network was able to obtain 6-fold lower RMSE than the conventional method, Varpro.

**Fig3** compares the performance of Varpro and the proposed network for in vivo $T_2$ mapping. The network's output retains more high-frequency features, e.g. in the cerebellum, and noise amplification is largely mitigated.

**Fig4** compares the performance of Varpro and the network to estimate $T_2^*$ maps, with similarly improved SNR and visualization of fine-scale detail, especially in the lower slices with poor $B_0$ inhomogeneity.

**Fig5** shows the CNN structure for MRF subspace coefficient estimation. On the right, we compared $T_1$ and $T_2$ maps synthesized by coefficient images and standard sliding-window reconstruction using the same 1-interleaf data, and compared with the 6-interleaf acquisition. Up to 1.4-fold RMSE improvement was obtained.

## Discussion

The proposed scan-specific network can leverage similarities across voxels during parameter estimation, and provides improved accuracy and robustness to noise than voxel-by-voxel estimation techniques. Incorporating Bloch modeling into the loss function allowed us to use deep learning libraries' automatic differentiation ability, which would lend itself to additional nonlinear models e.g. free-water mapping (10).

For MRF, we utilized a subspace approach to linearize the Bloch model while retaining high signal representation accuracy. This facilitated the optimization, and bypassed non-differentiable operators in dictionary-matching. Combining subspace modeling (11) with an unsupervised network allowed us to take advantage of temporal and spatial priors simultaneously.

A limitation is the training time (~12 hours/dataset). This can be improved using fine-tuning by pre-training the model, and rapidly refining in a few epochs on each individual subject's data.

## Conclusion

We demonstrated the ability of the proposed scan-specific, unsupervised network to improve the quantitative estimation accuracy over standard approaches. While using convolutional layers allowed for capitalizing on spatial relations between voxels during parameter estimation, utilizing a subspace approach for MRF facilitated the optimization with a differentiable loss function.

**Figures**

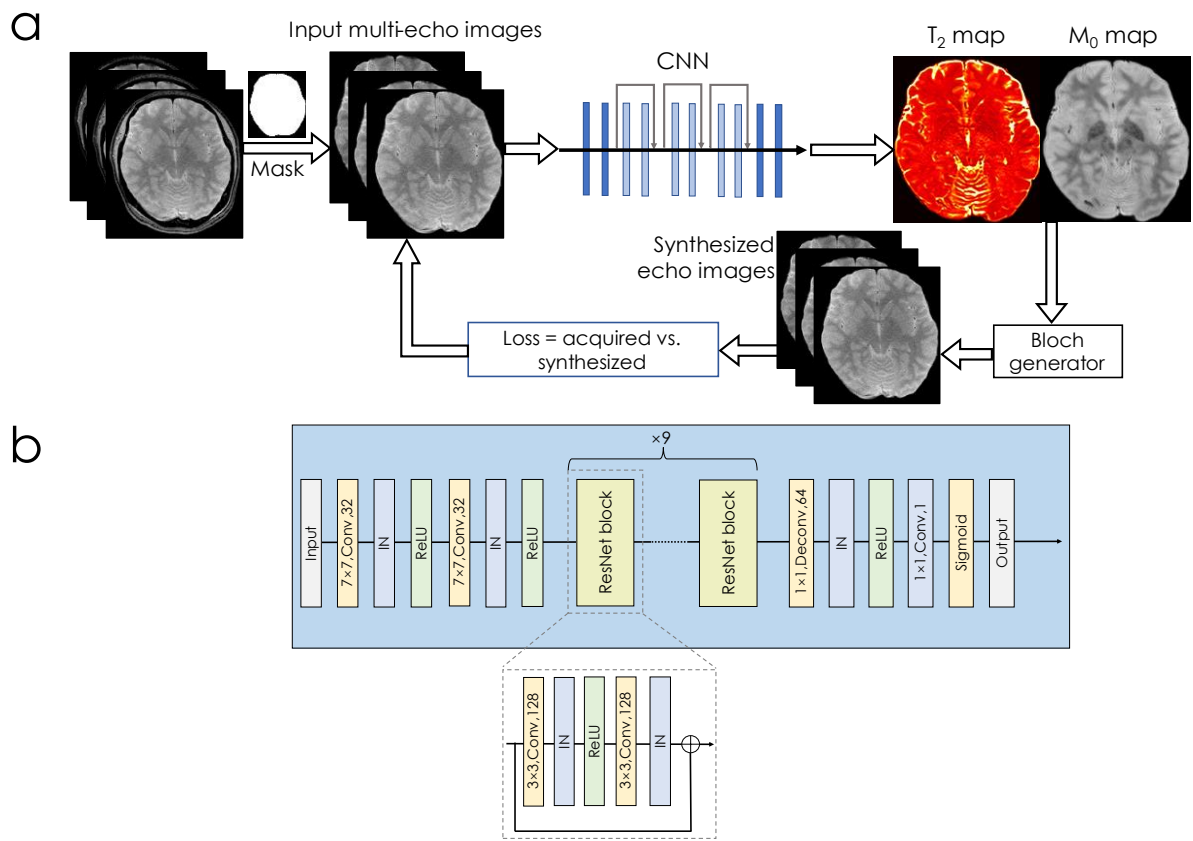

**Figure 1.** Proposed scan-specific unsupervised parameter mapping network. (a) Input multi-echo images are processed with the network to estimate $T_2$ and $M_0$ maps (for spin-echo imaging). These are then used for synthesizing echo images using Bloch modeling, which are compared against the input multi-contrast data in the loss function. (b) Detailed network structure, including 9 residual blocks.

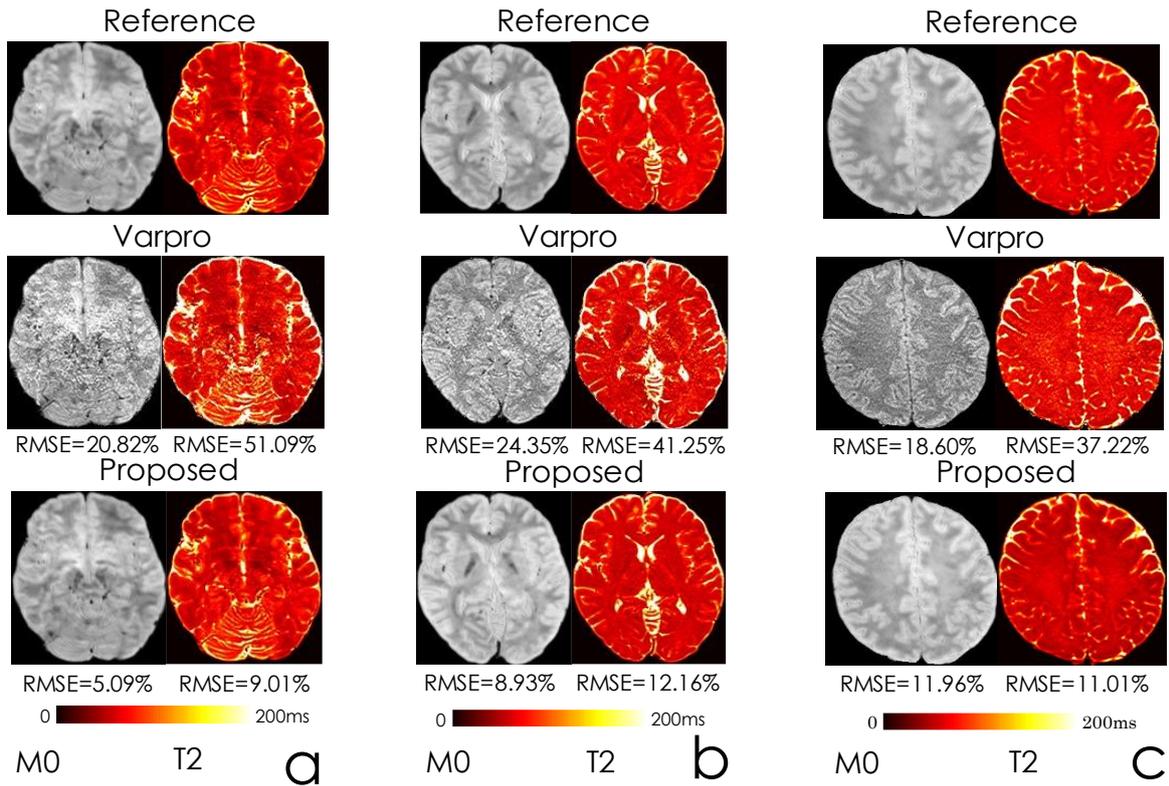

**Figure 2.** Three columns show different slices from numerical simulations, where Gaussian noise was added to input multi-echo images synthesized using the reference $M_0$ and $T_2$ maps in the top row. Varpro estimates the parameter maps on a voxel-by-voxel basis, thus suffering from noise propagation (middle row). Proposed unsupervised mapping provides up to a 6-fold reduction in RMSE by exploiting spatial relations through convolutions.

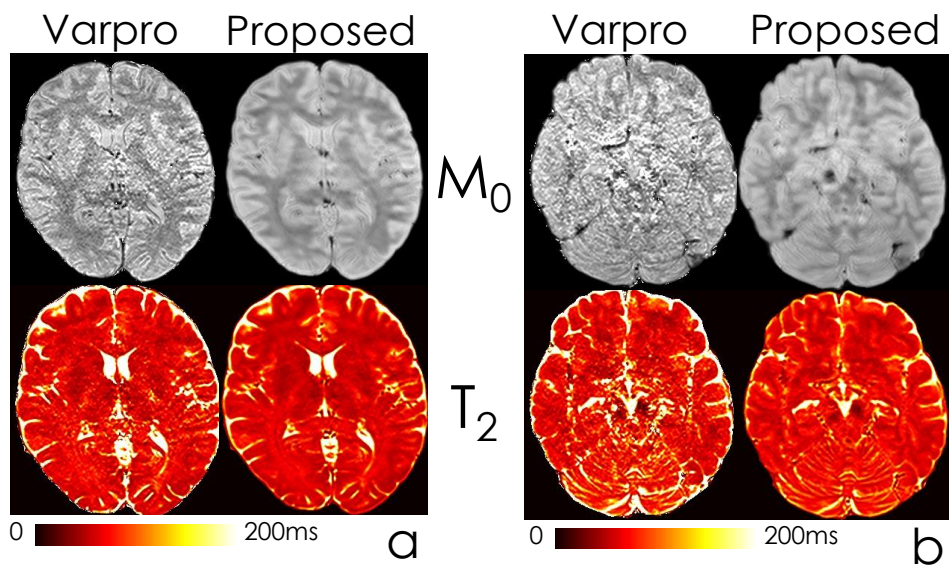

**Figure 3.** Compared with the traditional method (Varpro), the proposed scan-specific network is able to boost SNR and retain more fine-scale details in $T_2$ and $M_0$ map estimates from a set of spin-echo acquisitions.

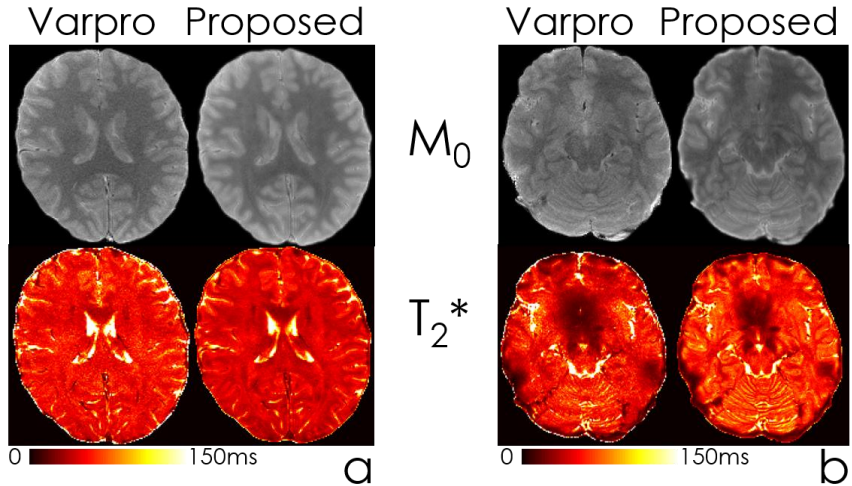

**Figure 4.** Compared with the traditional Varpro method, the network's reconstruction helps denoise $M_0$ and $T_2^*$ estimates, and yields fine details especially in the lower slice with increased B0 inhomogeneity, e.g. in the cerebellum.

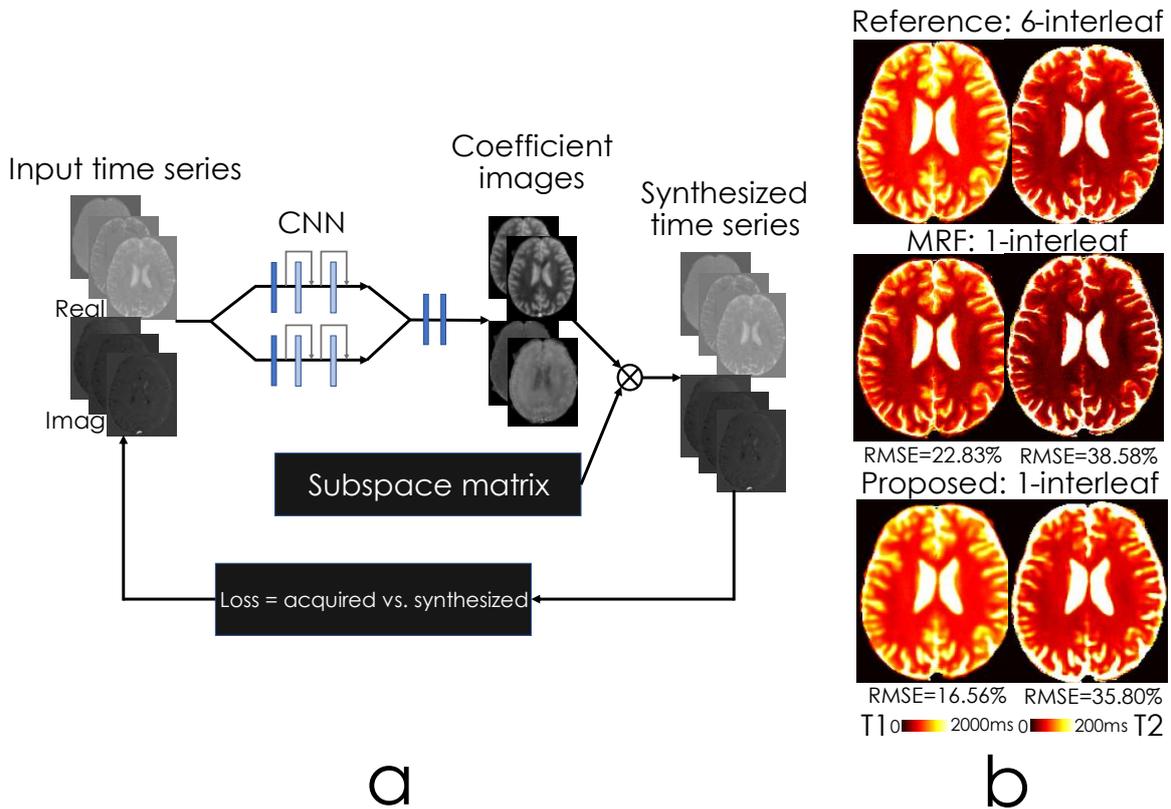

**Figure 5.** (a) Real and imaginary parts of MRF time series are presented as input to the network, which estimates subspace coefficients. Multiplication with the subspace matrix permits the synthesis of time series data, which are then compared to the input images in the loss function.
(b, top) Reference maps are estimated using sliding window reconstruction on 6-interleaf data. (middle) Maps from sliding window reconstruction on single-interleaf data. (bottom) Proposed unsupervised network's $T_1$, $T_2$ estimates based on single-interleaf input images.